\documentclass[aps, twocolumn]{revtex4-2}
\usepackage{bm}
\usepackage[margin=1.2in]{geometry}
\usepackage[utf8]{inputenc}
\usepackage{amsfonts}
\usepackage{amssymb}
\usepackage{graphicx}
\usepackage{amsmath}
\usepackage{ragged2e}
\usepackage{empheq}
\usepackage{tikz}
\usetikzlibrary{arrows,positioning,shapes,backgrounds,fit}  
\usetikzlibrary{tikzmark,calc}
\usepackage{caption}
\usepackage{float}
\DeclareSymbolFont{matha}{OML}{txmi}{m}{it}
\DeclareMathSymbol{\varv}{\mathord}{matha}{118}
\usepackage{soul}

\begin{document}

\begin{abstract}
As the simplest model of \textcolor{black}{transport of interacting particles} in a disordered medium, we consider the asymmetric simple exclusion process (ASEP) in which particles with hard-core interactions perform biased random walks, on the supercritical percolation cluster. In this process, the long time trajectory of a marked particle consists of steps on the backbone, punctuated by time spent in side-branches. We study the probability distribution in  the steady state of the waiting  time  $T_w$  of a randomly chosen particle, in a side-branch since its last step along the backbone. Exact numerical evaluation of this on a single side-branch of length $L=$ $1$  to $9$  shows that for large fields,  the probability distribution of $\log T_w $ has multiple well separated  peaks. We extend this result  to a  regular comb, and  to  \textcolor{black}{the} ASEP on  the percolation cluster. We show that  in the steady state, the fractional number of particles that have been in the same side-branch for a  time interval greater than $T_w$  varies as  $exp( - c \sqrt{\log T_w})$ for large $T_w$, where $c$ depends only on the bias field. However, these long timescales are not reflected in the eigenvalue spectrum of the Markov evolution matrix. The system shows dynamical heterogeneity, with particles segregating into pockets of high and low mobilities. 
\end{abstract}

\title{Asymmetric simple exclusion process on the percolation cluster: Waiting time distribution in side-branches}

\author{Chandrashekar Iyer and Mustansir Barma}
\affiliation{%
 Tata Institute of Fundamental Research, Gopanpally, Hyderabad, 500046, India}%

\author{Hunnervir Singh and Deepak Dhar}%
\affiliation{
Indian Institute of Science Education and Research, Dr. Homi Bhabha Road, Pashan, Pune 411008, India
}%

\maketitle

Fluid imbibition  and transport in disordered media like porous rocks have been topics of extensive study for many decades now, starting with the early experimental work of  Franklin \cite{franklin}, and  the mathematical model of percolation by Broadbent and Hammersley   \cite{broadbent-hammersley}.  The many applications in many fields, ranging from gel chromatography to cell transport to petroleum extraction,   have led  to  a very large amount of work  \cite{applications, devos, sahimi, biassed-diffusion-3}. 
Diffusion on the percolation network was studied in the paradigmatic model of `ant in a maze' by De Gennes \cite{de-gennes}.  Introduction of a  preferred direction in the diffusion process brings in new effects, both in the case of non-interacting particles \cite{biassed-diffusion-1, biassed-diffusion-2, biassed-diffusion-3}, and particles with on-site hard core interactions \cite{ramaswamy-barma}. The latter process, called the Asymmetric Simple Exclusion Process  (ASEP) has been studied extensively on regular lattices (without disorder) in the literature  \cite{derrida,mallick,liggett}.

For the case without interactions, the average drift velocity of a diffusing particle on a percolation network is a non-monotonic function of the field, and actually vanishes for all field strengths $E > E^*(p)$, \textcolor{black}{where $p$ is the fraction of bonds present \cite{barma-dhar}. \textcolor{black}{Here the driving field $E$  makes the steps of the walkers more likely along the field than against.} This is due to the trapping of the particle in side branches, and along backbends, which are \textcolor{black}{oppositely directed segments along the backbone connecting opposite sides of a sample.} For strong enough fields, the mean trapping time in a side branch diverges, and    the  large-time limit of the mean velocity becomes zero \cite{dhar-stauffer}.}

Hard-core interactions between particles gives rise to  two competing effects.  The particles get stuck in side branches, but this reduces the effective depth of the branch for the next incoming particle.  This decreases the  trapping time for the incoming particles. \textcolor{black}{However,}  particles that are buried in  a side branch below other trapped particles have a much increased trapping time.  The net result is that  the distribution of trapping times becomes very broad, with a very fat tail, but with finite mean. 

The time spent in the same side branch since the last step along the backbone will be called the \textcolor{black}{waiting} time in the side-branch. 
The question we ask here is: In the steady state, if we pick  at random a particle on the spanning cluster,  what is the probability that its waiting time is greater than $T_w$ in the current sidebranch after  the last step along the backbone?

We show that for large bias,  the answer is  non-trivial: the probability distribution of {\it  the logarithm of  $  T_w$  }  has multiple well-separated peaks. Each peak corresponds to particles at a particular depth from the backbone.  We argue that for the ASEP on the spanning cluster, the fractional number of particles which have remained in the same side branch for a time greater than $T_w$ decreases slower than any power law, as $\exp( -c\sqrt{\log T_w})$, where $c$ is a constant. \textcolor{black}{This form also describes the decay of velocity-velocity correlations.} Also, the steady state shows dynamical heterogeneity and many-body localization type behavior, with the emergence of high- and low-mobility pockets.

We consider  the percolation problem on a $d$-dimensional lattice, with a fraction $q = 1-p$ of bonds removed at random, with $p$ greater than the undirected percolation threshold $p_c$. We place hard-core particles at a fraction $\rho$ of the sites of the spanning cluster [Fig. 1]. \textcolor{black}{The sites of  the infinite cluster  that have at least two disjoint paths to the boundary of the lattice constitute the  backbone of the cluster.  The  remaining sites are side-branches.} The system  evolves in time by  Markovian evolution. \textcolor{black}{The system evolves in continuous time as follows: Each particle tries to jump to any of its unoccupied  nearest neighbors with a rate (1+$g$)/2 if the jump is in the direction of the field, and a  rate (1-$g$)/2 if it is against the field, independent of other particles. We assume periodic boundary conditions. In the computer implementation, we pick a bond of the spanning cluster at random, and change its state with probabilities shown in Fig. 2, and increase the clock time by 1/$B$, where $B$ is the number of bonds  in the spanning cluster.}

\begin{figure}
\begin{center}
\includegraphics[width= 7.5cm]{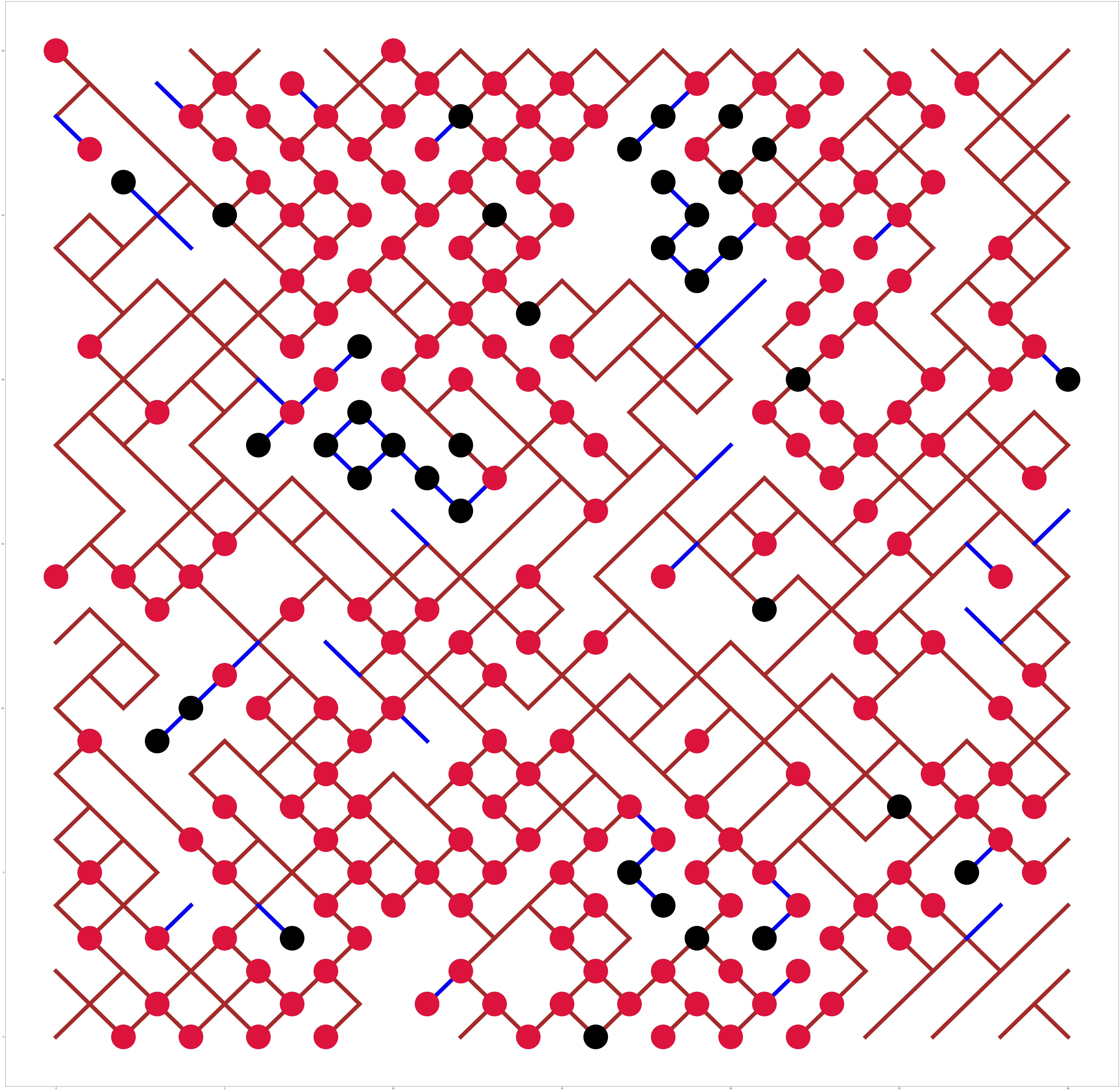}
\caption{ \justifying{A realization of the  spanning cluster  for a bond percolation. The bonds in the backbone and sidebranches  are colored brown and blue respectively. The particles are color coded red or black depending on whether their net displacement in the last $100$ time steps was of order $10^{2}$, or of order $1$.  Here $p = 0.67$, lattice size is $30 \times 30$, $\rho = 0.75$, and the bias parameter $g=0.75$.}}
\label{fig: startification}
\end{center}

\end{figure}

\begin{figure}
\begin{center}
\includegraphics[width=7.5cm]{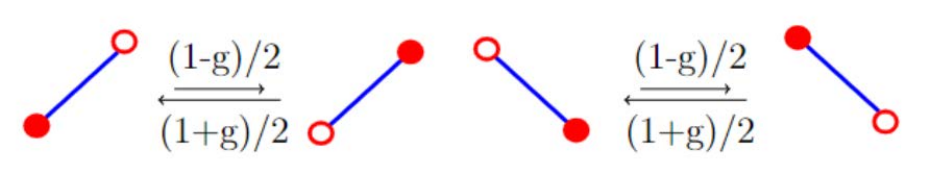}
\caption{ \justifying{\textcolor{black}{The rates of transition in the continuum time model.} Note that the actual jump takes place only if allowed by the hard core interactions.}}
\end{center}
\end{figure}
~\\
On the spanning cluster on an $L \times L$ lattice, with periodic boundary conditions, there is a unique long-time steady state.
In the steady state, \textcolor{black}{a tagged particle} moves along the backbone, then goes into a side branch, then comes out of it, and moves along the field, then get trapped again, and so on. The statistical properties of a tagged particle trajectory for long times are the same for all particles. Eventually, each particle  visits each site of the spanning cluster infinitely often, and 
spends a nonzero fraction of the total time at each of  the sites of the spanning cluster.

 The behavior of the system for small fields is quite easy to understand. For small bias $g$, the average current  density in the system is proportional \textcolor{black}{to $g$}. The coefficient of proportionality tends to zero, as $p$ tends to  $p_{c}$ \cite{ohtsuki}. 
 
The behavior of average current density  in the limit of infinitely strong fields ($g=1$) is also easy to understand. In this limit, the particle cannot move against the field. If $p$ is less than the directed percolation threshold $p_{d}$, there is no possibility of macroscopic transport \cite{supplemental}. 



In the following, we will discuss in more detail the current carrying phase,  for $g$ near $1$.

\begin{figure} [H]
\includegraphics[width=7cm]{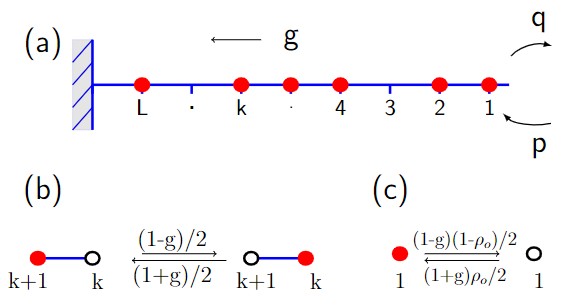}
\caption{\justifying{ (a) The ASEP process on a single sidebranch, open at one end. The rates of transitions between configurations with changes in the bulk shown in (b), and at the boundary site in (c).}}
\end{figure}

 We  first consider the ASEP process on a single line  of $L$ sites, closed at the left end, and open at the right end (Fig 3). At each site $i$, there is a binary variable $n_i$ taking value \textcolor{black}{$1$ or $0$}, depending on whether \textcolor{black}{or not} there is a particle at that site.  The particles can enter or leave the line from the open end at $i=1$. The site $i=0$ is an external site to the branch,  \textcolor{black}{from where one} can  introduce, or take out particles from the line.   On the line the evolution follows asymmetric exclusion process rules. We consider discrete time evolution: at each time step, we pick one of the bonds $(i,i+1)$ to update, with $i\in [0,L-1]$,  with equal probability.  If  $n_i  =1$, we exchange $n_i$ and $n_{i+1}$ with probability $(1+g)/2$.  If $n_i =0$, we exchange $n_i$ and $n_{i+1}$ with probability $(1-g)/2$. If the occupation numbers of the two sites are the same, the exchange will not change the configuration. If the picked value of $i$ is $0$, we first assign an occupation variable $n_0$ to the site $i=0$  independent of history: it is occupied with probability $\rho_o$ and unoccupied with probability $(1-\rho_0)$.   

It is easily verified that the exact steady state probability distribution is a product measure : 
Each site is independently occupied, with the activity $z_d$  at a site with depth $d$ being $ z_0 \left( \frac{1+g}{1-g}\right)^d$, with $z_0 = \rho_0/(1 -\rho_0)$.

For the discrete time evolution, we will take  $\Tilde{W} = 1 + W/ B$, where $W$ is the transition matrix in the continuous time formalism and $B$ is the number of bonds in the spanning cluster. Then the probability vector evolves by $|P(t + 1/B) \rangle = \Tilde{W} |P(t)\rangle.$ Consider a sub-matrix $\Tilde{W_j}$ \textcolor{black}{consisting} only of configurations that have \textcolor{black}{$j$ or more} particles. \textcolor{black}{$\Tilde{W_{j}}$ is simply derived from $\Tilde{W}$ by deleting the rows and columns corresponding to configurations consisting of less than $j$ particles \cite{supplemental}}.
Then, if we start with a configuration $|C\rangle$ of \textcolor{black}{$j$ particles}, the probability that the system has remained in this subspace \textcolor{black}{upto $T_w$ microsteps} is given by  

\begin{equation}
{\rm Prob}( {\rm Waiting~time } > T_w) = \sum_{C'} \langle C'| [\Tilde{W_j}]^{B T_w}|C\rangle.
\end{equation}

If we take $C$ according to the steady state measure,  but with the restriction that the number of particles in $C$ is exactly $j$, and the rightmost particle has just entered the chain and is at \textcolor{black}{site} $1$, this exactly gives the probability that the  time  spent in the side-branch is greater than $T_{w}$.

Finally,  knowing  the matrix $\Tilde{W}^{2^{n}}$ \cite{supplemental},  the matrix $\Tilde{W}^{2^{n+1}}$ can be obtained by a single matrix multiplication by squaring. Thus, we can efficiently calculate exactly the probability that the waiting time of the $j$-th particle in the branch lies between $2^n$ and $2^{n+1}$, for all integer $n$. Using this for different $j$, we can determine the probability that a particle has not left the system up to time $T_w$ (Fig. 4) \cite{supplemental}.

\textcolor{black}{Explicit implementation of  this procedure results in the probability distribution of trapping times are shown in Fig. 4.  Note that the wide range of values of $\log_2 T_w$. The results do not have any Monte Carlo errors, only roundoff errors that we ensured are neglible.}

\begin{figure}
\includegraphics[width = 8cm]{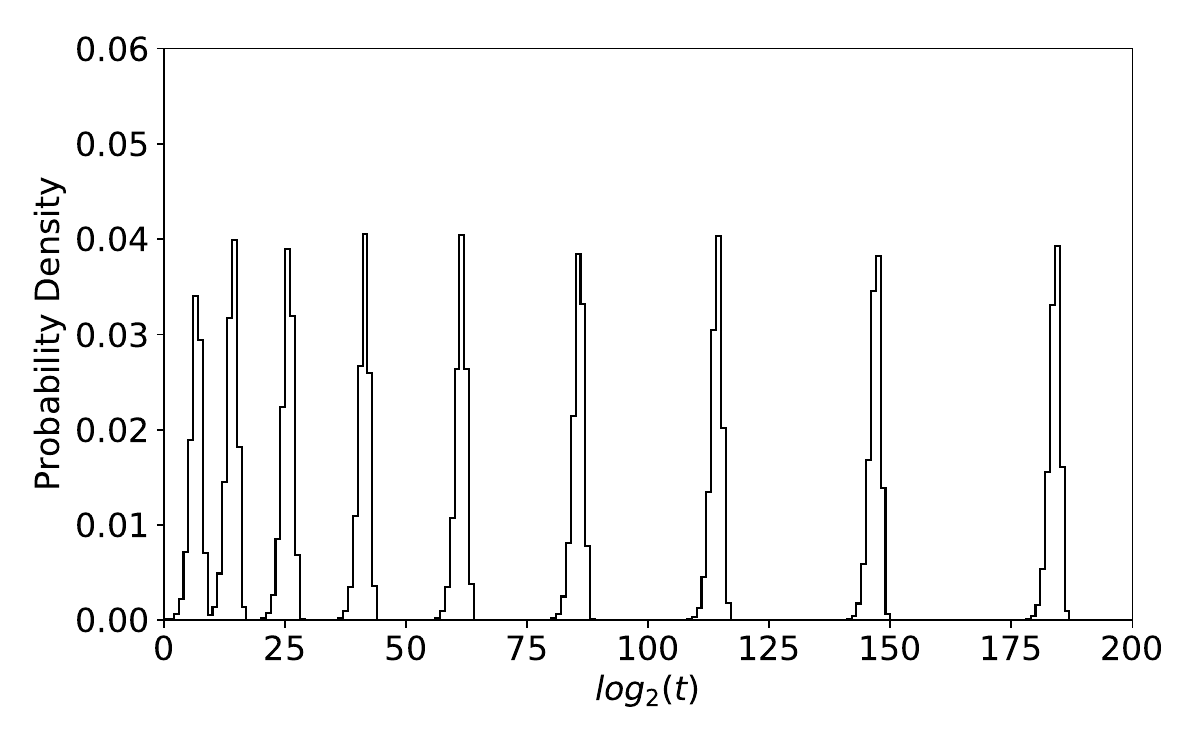}
\caption{ \justifying{Histogram showing the exact probability distribution in the steady state of a single side branch  of length $L=9$. Here, a randomly chosen particle has already spent a time $T_w$ in the side branch, with  $ n \leq \log_2 T_w\leq (n+1)$ for integers $n$ in the x-axis. The case shown is  for bias field $g=0.9$, and the parameter $z_0 =0.3$. The asymmetric shapes of the peaks is due to the centers of peaks not being at the mid-point of the bins.}}

\end{figure}

Fig. 5 shows the corresponding plot for the  cumulative probability that the logarithm of  waiting time is greater than  $T_w$, as a function of $\sqrt{\log (T_w)}$. The distribution \textcolor{black}{is} well approximated as a number of equal-sized steps.  For large bias, the probability that a \textcolor{black}{site} at depth $d$ is \textcolor{black}{occupied is close to unity} for $1\leq d \leq L$. The probability that the particle will remain in the branch upto time $T_w$ may be approximated as $\exp( - T_w /\tau_d)$, with $\tau_d =  C \left[\frac{1+g}{1-g}\right]^{  d(d+1)/2}$, as the system has an entropic bottleneck, \textcolor{black}{and the particle at site $i$ cannot exit until} there are no particles to the right of $i=d+1$. Here $C$ is a constant that depends on the density $\rho$.  As a function of $\log T_{w}$, this is almost a step function. We sum over $d$, the probability that the selected particle is at depth $d$, and the conditional probability that it survives for a time $>T_{w}$ given that it is at depth $d$, we get the approximate formula

\begin{eqnarray} 
     {\rm Prob}( {\rm Waiting~time} > T_w)  \\ \nonumber \MoveEqLeft \sim  1 - \frac{1}{L} \Bigg \lfloor \left[ \sqrt{   \frac{2 \log (T_w/C)}{ \log [(1+g)/(1-g)]}}\right] \Bigg \rfloor 
\end{eqnarray}

In Fig. 5, we have \textcolor{black}{also} shown a fit to the numerically determined cumulative function using a single fitting parameter $C$.

\begin{figure} [H]
\begin{center}
\includegraphics[width= 7.5 cm, height=4.5 cm]{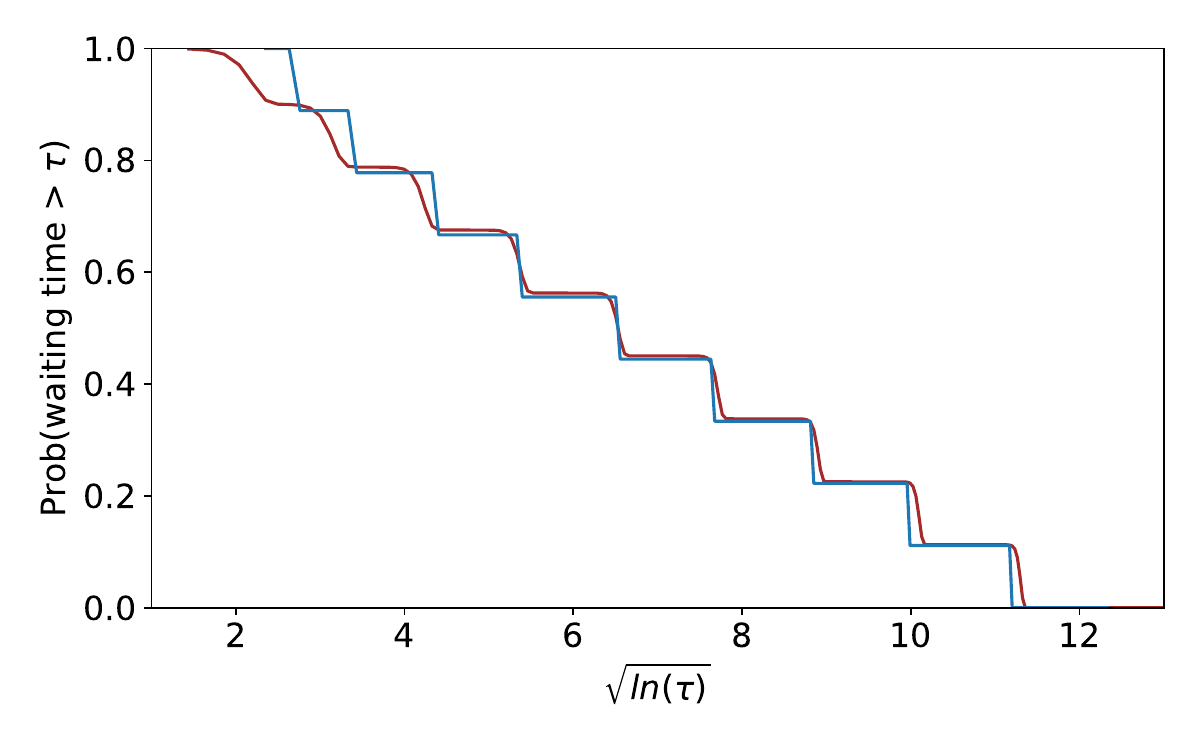}
\caption{ \justifying{The plot shows the probability of waiting time greater than $\tau$ as a function of $\log \tau$. The blue curve is a fit to the approximate extrapolation form (Eq. 2) using a single free parameter $C = 200$.}}
\end{center}
\end{figure}

Note that the eigenvalue spectrum of the full transition matrix $W$ corresponding to the original Markov process has a non-zero gap even in the limit of large $L$, and shows no indication of \textcolor{black}{a} diverging  time scale. The waiting time distribution  is a  persistence property, \textcolor{black}{which} can show  nontrivial behavior even for systems with simple $n$-point correlation functions \cite{persistence}, where some configurations occur with extremely low probabilities in the steady state. One finds signatures of slow decay in the corresponding eigenfunction and also in the largest eigenvalue of the restricted $W$-matrices, namely $\Tilde{W_{j}}$, as defined in Eq. 1 above \cite{supplemental}.

~\\


Now, let us consider a regular comb graph, in which there is a ring of $M$ sites, and a linear side-branch of depth $L$ attached to each site on this ring, with $M>>L$.  It is easily verified that the steady state on this graph is given by the  product measure. As the evolution within  a side-branch is affected only through the entry and exit of particles, we would expect that the behavior of waiting times in the ASEP on the regular comb would be qualitatively similar to that on a single branch.  

In Fig. 7, we have compared the result of  distribution of waiting times as obtained from Monte Carlo simulation of a regular comb with $M$ =100, $L$=4, with the calculation for the same value of $z_{0}$ and $g$ on a single branch.  We see that the agreement is excellent. This also suggests a further generalization to the problem of ASEP on a  random comb graph  (Fig. 6).


\begin{figure} [H]
\begin{center}
\includegraphics[width=3.5cm, height=4 cm] {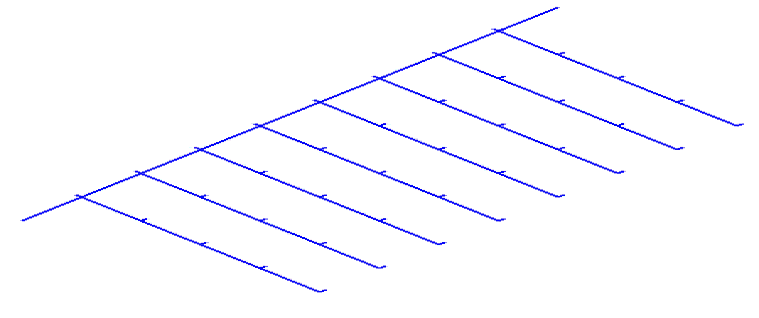}
\includegraphics[width=3.5cm, height=4 cm] {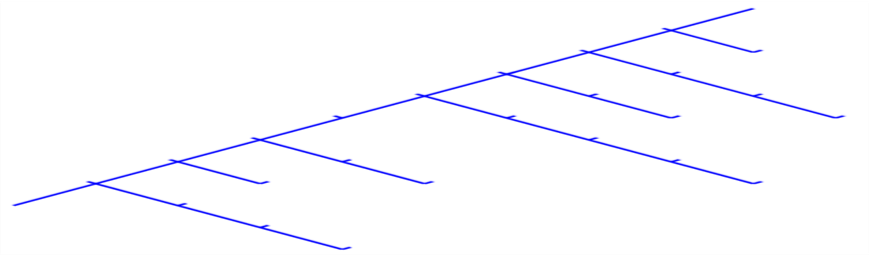}
\caption{ \justifying{Left: A regular comb graph. Right: A random comb graph. Different branches of varying lengths are attached to a single backbone line in a random comb, while the lengths are same in a regular comb. The direction of the field is downwards.}}
\end{center}
\end{figure}

\begin{figure} [H]
\begin{center}
\includegraphics[width=7.5cm, height=5 cm]{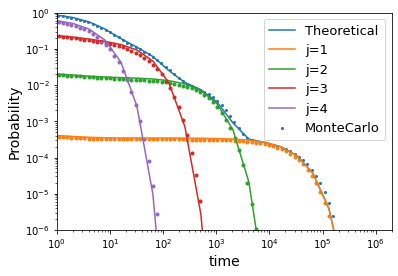}
\caption{ \justifying{The waiting time distributions obtained from the theoretical calculation for a single branch and compared to Monte Carlo simulations, for a system of Regular comb of ring length $M=100$, and branch depth $L = 4$, with a bias of $g=0.50$.}}
\end{center}
\end{figure}


For the random comb model \cite{random-comb-1, random-comb-2, random-comb-3, random-comb-4, random-comb-5, random-comb-6, random-comb-7}(Fig. 6), we choose the distribution of branch lengths to be the same as in the percolation problem, where the probability of a dead-end branch of length $d$ \textcolor{black}{decreases exponentially with $d$}, as $exp(-d/\xi_p)$, $\xi_p$ being the percolation correlation length. Then, the probability that the waiting time is greater than $T_{w}$ is approximately equal to the fractional number of sites at depth greater than $d$ such that $ \tau_d > T_{w}$. In the percolation problem, the fractional number of sites at distance $d$ from the backbone $\sim \exp( -d/\xi_p)$.
 This immediately gives, for the percolation problem,
\begin{equation}
{\rm Prob( Waiting ~time} > T_w) \sim \exp\left( -c \sqrt{ \log (T_w/t_{o})}\right), 
\end{equation}
which is a slower decay than any power law. The behavior holds for all non-zero bias, but \textcolor{black}{for smaller bias values} the asymptotic behavior will be seen only for larger $T_{w}$. There is a variation of the density of along the backbone, and the parameter $t_o$ will vary a bit. But this does not change the  asymptotic behavior.  \textcolor{black}{On a regular hypercubical lattice of linear extent $M$,  the largest trapping time is of order $M^{c (log M)}$, in contrast to the non-interacting case, where the time grows as  $M^y$ \cite{random-comb-1}, where $c$ and $y$ are field dependent constants.}

\textcolor{black}{We have verified the bias-dependence of waiting time distributions on a two-dimensional supercritical percolation cluster by showing a scaling collapse for multiple bias values in the End Matter.}

The fact that trapping times of particles at different depths can differ from each other by orders of magnitude at strong fields has an interesting consequence \textcolor{black}{for} the average displacement of particles  in a time interval $T$ in steady state.  For any choice of $T$, there \textcolor{black}{are} particles that are essentially stuck, and show effectively no displacement, while particles at depths $d$ corresponding to smaller trapping times \textcolor{black}{are} mobile. If the time scale $T$ lies in the gap between the trapping times of two adjacent values of $d$, the population of particles \textcolor{black}{may be considered as divided into essentially separate} mobile and immobile classes.

 Separation of particles into more - and less - active regions has been discussed in the context of dynamical heterogeneity in structural glasses \cite{heterogeneity1,heterogeneity2}. Here, the origin of the heterogeneity is geometrical, and remains quenched in space. \textcolor{black}{A similar behaviour is found in disordered quantum systems of interacting particles which display many body localization} \cite{many-body-localization-1,many-body-localization-2}.  Very large trapping times for deeply-buried grains have been discussed earlier in models of sandpiles and ricepiles \cite{pradhan-dd}. \textcolor{black}{In the End Matter, we argue that the velocity distribution of the interacting particles on a two-dimensional supercritical percolation cluster also shows a slow decay of the form $exp(-\sqrt{log t})$.}

To summarize, in this paper,  we have shown that the distribution of waiting times $T_{w}$ in the ASEP on percolation clusters in strong fields is very broad.  Also, the distribution of $\log T_{w}$ has multiple well separated peaks. For any time scale of observation $T_{w}$, there is a significant fraction of particles that are deeply buried and localized. This fraction is quite robust, and varies very slowly with $T_{w}$. We have argued that it varies as $exp( -c \sqrt{\log T_{w}})$,  for large $T_{w}$, which is slower than any power law. We emphasize that this is a typical behavior, and not a large deviation property.

\paragraph*{\it Acknowledgements}C.I. and D.D. thank TIFR Hyderabad for hospitality. M.B. and D.D. acknowledge the support of the Indian National Science Academy(INSA). \textcolor{black}{We acknowledge the support of the Department of Atomic Energy, Government of India, under Project Identification No. RTI4007.}


\newpage

\begin{center}
\section*{End Matter}
\end{center}

\textit{Appendix A: Verifying bias-dependence of waiting times on a percolation cluster --}

\begin{figure} [H]
\begin{center}
\includegraphics[width=8cm, height=5 cm]{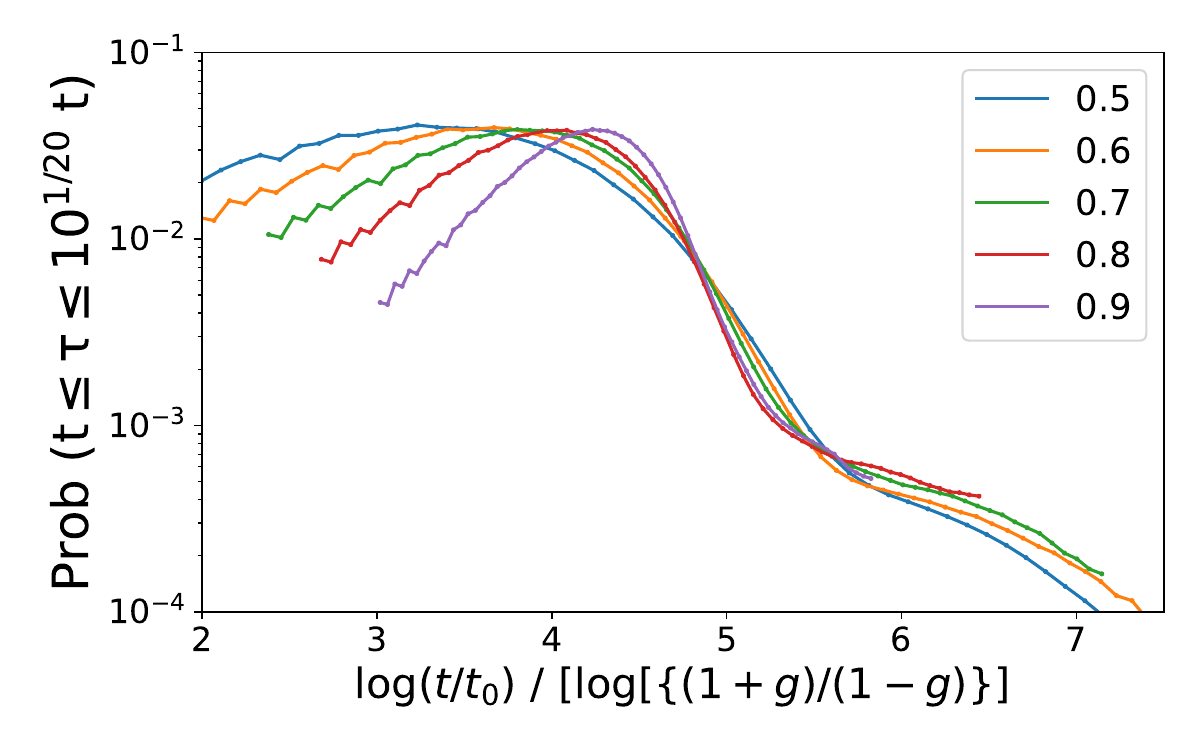}
\caption{ \justifying{\textcolor{black}{Plot of  waiting time $\tau$ falling in the bin $[t, 10^{1/20}t]$ versus $log (t/t_{0})/ \log\Big[(1+g)/(1-g)\Big]$. This scaled distribution of waiting times is obtained through Monte Carlo simulations on a percolation cluster of size $60 \times 60$, with $\rho = 0.5$. Here $p = 0.67$, and the maximum time of numerical simulation is $10^{5}$ MCS, averaged over 128 trajectories for each bias.}}}
\end{center}
\end{figure}

\textcolor{black}{The probability that the waiting time is greater than $t$ varies as the fraction of sites at a depth $>$ $D(t)$, where the trapping time of particles at depth $D(t)$ is of order $t$, i.e.  $\Big[\frac{(1+g)}{(1-g)}\Big]^{D^2/2} \approx t/t_{o}$.  This implies that if we plot the waiting time distribution versus  $\log (t/t_{o})/ \log \Big[\frac{(1+g)}{(1-g)} \Big]$,  curves for different values of $g$ will fall nearly on top of each other. This is verified in Fig. 8, where we see a scaling collapse for different values of the bias, when the distribution function is plotted against $log (t/t_{0})/ \log\Big[\frac{(1+g)}{(1-g)}\Big]$, where $t_{0}$ \textcolor{black}{is a bias-dependent fitting parameter that plays the role of the parameter  $C$  used in Eq. 2, in the discussion  of a single branch}. We are not able to explicitly verify the asymptotic  $exp(-\sqrt{\log t})$ dependence from the simulations of total duration of only order $10^5$ MCS,  compared to the timescales of order $10^{60}$   reachable using exact matrix diagonalization (Figs. 4 and 5).}   

\paragraph*{}
\paragraph*{\textit{Appendix B: Velocity distribution of particles and velocity-velocity correlations --}} 

\paragraph*{}\textcolor{black}{Consider a particular realization of the evolution of the system. We define the average velocity of a marked grain   in the time window  $[t -T/2, t+T/2]$ as $(\vec{X}(t+T/2) -\vec{X}(t-T/2))/T$, and denote it by $v(t,T)$.  This is a random variable, and  distribution  of the logarithm of the $v(t,T)$  is shown in Fig. 9. There is a small probability of $v$ taking negative values for $T>>1$, and these are omitted from the graph, but a significant fraction show velocity exactly zero.  These are due to deeply buried particles, that did not manage to come out in the period of observation.  These are displayed in the leftmost bin in the histogram.  We see that for large $T$, the values span many orders of magnitude.  For large $T$, there is a peak at a value near $\langle v \rangle$, where  $\langle v \rangle$ is the ensemble average, and this peak becomes sharper for larger $T$, as expected by the law of large numbers. Also, the size of the left peak decreases as $T$ increases.} \textcolor{black}{This also implies that the variance of $[X(t +\tau)-X(t)]$ will increase  approximately as $\tau^2$ for large $\tau$.}

\begin{figure} [h]
\begin{center}
\includegraphics[width=8cm, height=5cm]{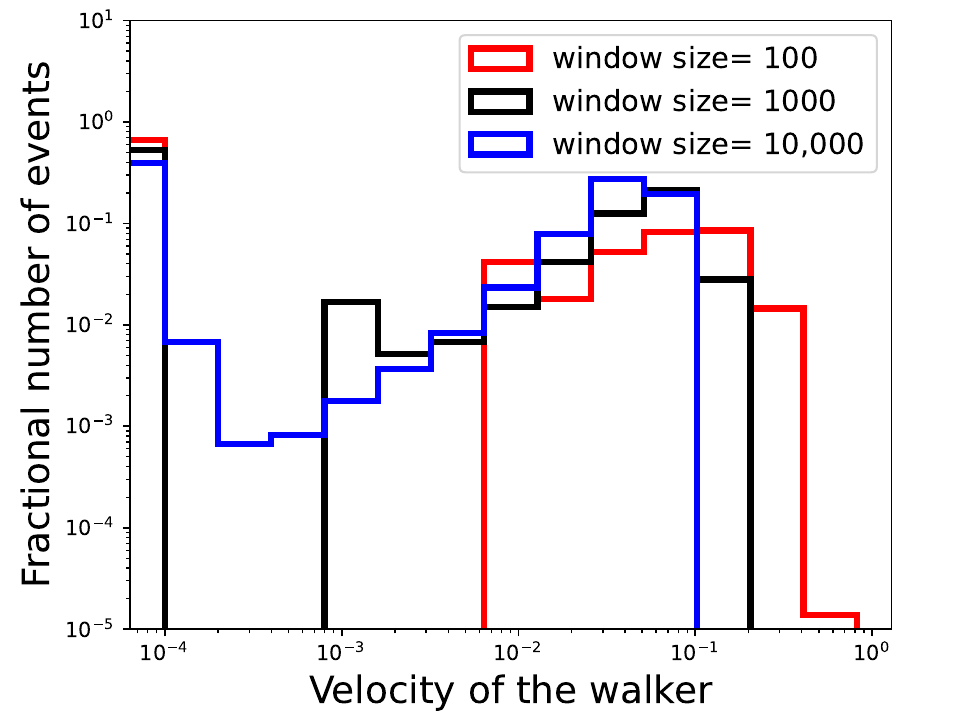}
\caption{ \justifying{Histogram of probabilities of different values of velocities obtained in simulation on a regular comb of size $M$=100, $L$=5, with density of particles =0.8.  Averaging is done  over the trajectories of all particles for a single history of the configuration of duration $5 \times 10^6$.  The leftmost bin contains all entries with value  $< 10^{-4}$, including zero and negative values. All other bins are of width log 2.}}
\end{center}
\end{figure}

\textcolor{black}{Now let us consider the  velocity-velocity auto-correlation function  of a marked grain in the steady state of the system. Denote by $F(\tau)$  the connected part of $\langle v(t,T)v(t+\tau,T) \rangle$. Even for large values of $T$,  there will be some deeply buried grains for which $v(t,T)$ and $v(t+\tau, T)$ are both exactly zero.  The fractional number of these decreases as  $exp( -c\sqrt{\log \tau})$, for $\tau >>T$.  These give rise to a positive contribution to the autocorrelation function that varies as $exp(-c\sqrt{\log \tau})$, for $\tau >> T$. We note that this is much slower than the $\tau^{-D/2}$ decay, which comes from hydrodynamical effects in $D$ dimensions in systems without disorder \cite{supplemental}.} 

\textcolor{black}{In the SM, we show some Monte Carlo data about the velocity-velocity correlation function on a regular comb. The behavior is in a satisfactory agreement   with the theory \cite{supplemental}.}  

\paragraph*{}
\paragraph*{\textit{Appendix C: Long time scales not seen in the eigenvalues of the relaxation matrix--}}


\paragraph*{}Consider a general discrete time Markov chain on a discrete state space. Let the probability of configuration $C$ in the steady state be denoted by $Prob(C)$. It is possible that some configurations have very low but non-zero probability of being in the steady state. Let $F$ be such a configuration, with $Prob(F) = \epsilon << 1$.

Then, in a long time series of the evolution of system of length $T$, the expected number of occurrences of $F$ will be $\epsilon T$, and the average time interval between two occurrences of $F$ will be $1/\epsilon >>1$. \textit{This time scale is set by the magnitude of small valued entries in the steady state vector, and not by the spectrum of relaxation matrix.}

As a very simple example to illustrate this point, consider time evolution of $N$ independent Ising spins, each flipping in continuous time, independently of other spins, with rate 1. Then, it is easy to see that the relaxation matrix is $W = \sum ( \sigma^{x}_{i} - 1)$, and the eigenvalues are negative integers. \textcolor{black}{More generally, in any discrete Markov chain evolution, the average time between revisits of any subset S of configurations is the inverse probability of that set}. However, the time between consecutive occurrences of any configuration is of order $2^{N}$, the inverse of which is not seen in the eigenvalue spectrum of $W$. In our problem of calculating probability of residence times of traps, the `entropic bottlenecks’ of configurations are where all sites to right of site $d$ are empty gives rise to large residence times.

The explanation as to why  the very long  residence times are not captured in the eigenvalues of the relaxation matrix is this :  The residence time is not an observable which is function of  the instantaneous configuration of indistinguishable  particles.  In order to define it, we have to tag at least one  particle, and that makes the configuration space much larger. There will be   some eigenvalues  of  the  much  higher-dimensional  transition matrix of the system with one tagged particle that are much closer to $1$.

\newpage

\onecolumngrid

\setcounter{figure}{0}
\renewcommand{\thefigure}{S\arabic{figure}}

\setcounter{equation}{0}
\def\theequation{S\arabic{equation}}
~\\
~\\
~\\
{\huge Supplemental Material: Asymmetric simple exclusion process on the percolation cluster: Waiting time distribution in side-branches}\\

\subsection*{A. Calculation of residence and waiting times}

Let $|\psi\rangle$ be the steady state  eigenvector for the full matrix $\Tilde{W}$. We define  projection operators 
$P_{y,j}$  that project the vector to the subspace of configurations in which the site at depth $y$ is occupied, by the $j$-th
particle counting from below.  Then, summing over allowed values of $j$, we get the projection operator $P_{y}$, which projects onto the subspace in which site at depth $y$ is occupied. 

Then  $\langle1111..|P_{y,j} |\psi\rangle$ gives the probability that site $y$ is occupied in the steady state, by the $j$-th particle from below, and $\langle1111..|   \Tilde{W_{j}}^{T.B} P_{y,j} |\psi\rangle$  gives the   probability that the particle remains in the side-branch in the entire interval $[0,T]$.  Summing this over $j$, we get the probability that  there was a particle at  site $y$, and it remained in the same branch till time $T$. If $y=1$, this gives the probability that  residence time in the branch is $>T$  for a  particle that has just entered the branch. 

\subsection*{B. Implementing the algorithm: Example and technical details}

In this section, we work out a simple example of finding the waiting time probability using the transition matrix. 

We take a branch of length $L=2$, and describe the system as shown in Figure 1 below. There are two sites in the branch, and hence, 4 configurations: $|00\rangle, |01\rangle, |10\rangle$ and $|11\rangle$. There are transitions from one configuration to another, with the rates $p, q, p^{\prime}, q^{\prime}$ as shown. In the context of the current problem, the rates $p=(1+g)\rho_{o}/2$, $q=(1-g)(1-\rho_{o})/2$, $p^{\prime}=(1+g)/2$ and $q^{\prime}=(1-g)/2$.

\begin{figure}[h] 
    \centering   \includegraphics [width=13cm, height=5cm]{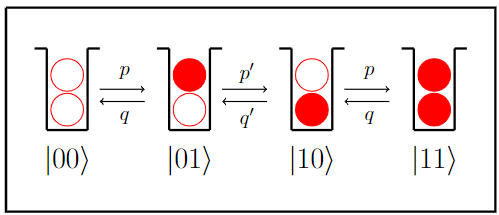}       
    \caption{\it Schematic showing configurations and the rates of transitions between them.} 

\end{figure}


For the above system, we obtain the transition matrix $W$ as follows:

\begin{equation*}
    W = \begin{pmatrix}
-p & q & 0 & 0\\
p & -(q + p^{\prime})& q^{\prime} & 0\\
0 & p^{\prime} & -(p + q^{\prime}) & q\\
0 & 0 & p & -q
\end{pmatrix} \tag{S1} 
\end{equation*}


Note that the matrix has dimensions of 4 $\times$ 4. For larger $L$, the dimensions increase as $2^{L} \times 2^{L}$. From Eq. (1) in the main text, we need to solve for all $1 \geq j \geq 2$, where $j$ is the number of particles in the branch including the last recently added particle. We explicitly show the calculation below. 

\subsubsection*{\underline{$j=1$}}

In this case, we need to include only those configurations which have $j=1$ particles or above, viz. $|01\rangle, |10\rangle$ and $|11\rangle$. In this case, the restricted matrix $W_{j} \equiv W_{1}$ becomes: 

\begin{equation*}
   W_{1} = \begin{pmatrix}
-(q + p^{\prime})& q^{\prime} & 0\\
p^{\prime} & -(p + q^{\prime}) & q\\
0 & p & -q
\end{pmatrix} ; |C\rangle = \begin{pmatrix}
   P_{ss}(|01\rangle) \\ 0 \\ 0 \\ 
\end{pmatrix} \tag{S2}
\end{equation*}


\begin{flushleft}
where, $P_{ss}(|01\rangle)$ is the steady state probability of the configuration $|01\rangle$. 
\end{flushleft}

One may work out the steady state weights relative to the configuration $|00\rangle$. For $|01\rangle$, the weight is $w_{2}$, for $|10\rangle$ it is $(w_{1}w_{2})$ and for $|11\rangle$ it is $(w_{1}w_{2}^{2})$, where $w_{1} = (1+g)/(1-g)$ and $w_{2} = [(1+g)\rho_{o}]/[(1-g)(1-\rho_{o})]$. Note that these values in $|C\rangle$ must be normalised. 

For the discrete time evolution, we have used the matrix $\Tilde{W}$, defined on page 3 in the main text. Observe that the value in place of $P_{ss}(|11\rangle)$ and $P_{ss}(|10\rangle)$ will be 0. This is because $|11\rangle$ is a configuration consisting of $2$ particles, and we had started with the initial condition that there must be only $j=1$ particles in the system. The weight of $|10\rangle$ is zero as well, because one particle must necessarily be present at the first site as the initial condition. In this case, $|C\rangle^{T} = (1, 0, 0)$. 

 \subsubsection*{\underline{$j=2$}}

Here, it is straightforward to see that $W_{2}$ is a $1 \times 1$ matrix, corresponding only to the configuration $|11\rangle$. $W_{2} = (-q)$, $|C\rangle = 1 |11\rangle$.

Figure 2 shows the plot for $L=1$ as well. $L=1$ has simply 1 term in the computation, and it is the same as that computed for $j=2$ in the above example.

 \begin{figure}[h]
    \centering   \includegraphics [width=13cm, height=8cm]{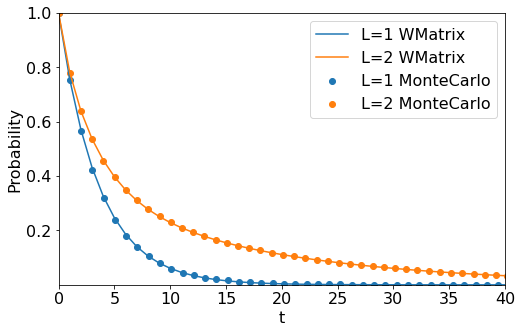}
    \caption{\it Plot showing the waiting time probability, Prob(waiting time > t), calculated for L=1 and L=2, with the Monte Carlo agreement plotted together.}
    \label{fig:enter-label}
\end{figure}


\textcolor{black}{For larger systems, we evaluate the probability at time $t$ in powers of 2, i.e. $t= 2^{n}$, for $n=$ 1 to 200.  This can be done efficiently, using $\Tilde{W}^{2^{n+1}} = \Tilde{W}^{2^{n}} . \Tilde{W}^{2^{n}}$. This gives us the exact binned probabilities that the survival/trapping time lies between $2^{n}$ and $2^{n+1}$. This allows us to draw the exact numerically determined histogram in Fig 2. The errors are controlled, and do not exceed $10^{-8}$.}

\subsubsection*{Technical details}

We have implemented the program using appropriate python libraries. For purposes of storing the W-Matrix, “scipy.sparse” was used. "scipy.linalg” was used to implement matrix multiplication, and “mpmath” library was used to implement arbitrary precision calculations, to prevent overflow/underflow errors.

\subsubsection*{Eigenvalues of the transition matrices}

Here, we emphasize the point that the eigenvalues of the transition matrix $W$ by itself do not show any signatures of the large timescales mentioned in the main text. 

Putting the values of $g$ and $\rho$ as $0.9$ and $0.3$ respectively, we get the eigenvalues of the probability matrix obtained from $\Tilde{W}$ for $L=2$ as: 0.344926, 0.726111, 0.774005 and 1, giving the largest timescale $\sim 4$. However, the restricted probability matrix of $\Tilde{W}_{1}$ has eigenvalues: 0.3489919, 0.7388199 and 0.999807, giving the largest timescale $\sim 5000$.

\subsection*{C. The velocity-velocity autocorrelation function} 

Here we give some details of the argument for the long-time dependence of the velocity-velocity autocorrelation function $\langle v(t) v(t +\tau) \rangle_{c}$.

We note that there is a non-zero probability that a marked particle is stuck in a trap of trapping
time greater than $\tau$, and the value of $v(t)$ and $v(t+\tau)$ are both exactly zero. \textcolor{red}{Since the mean velocity <v> is non-zero positive, this implies that these events give a positive contribution to the velocity-velocity auto-correlation function, whose value , for large $\tau$ decreases with $\tau$ at the same rate as the probability that the waiting time is greater than $\tau$.} 

The velocity-velocity autocorrelation function has a maximum at a finite $\tau$, (and even becomes  negative for $\tau \sim 1$), and for small $\tau$, there is a significant probability that the next step will be the reverse of the previous step, ( similar to rattling motion in a cage). At large time lags, the dominant part in the velocity-velocity autocorrelation is due to the trapped walkers. As the fraction of the trapped walkers decreases on increasing $\tau$, the autocorrelation function also decreases and shows bumps corresponding to the trapping at different depths. 

\textcolor{black}{Thus, we conclude that in the ASEP on the percolation cluster, the velocity-velocity autocorrelation will decay as $\exp(-\sqrt {\log \tau})$ for large $\tau$. In fig 3 below, we show the function $\langle v(t)v(t + \tau)\rangle_{c}$ versus $\log \tau$ for a regular comb  with $L=100$, and branches of depth 5, and density =0.8. The averages were calculated using the trajectories of all particles, using a single run of duration 5 $\times$ $10^{6}$ MCS. We clearly see that the function decays slowly with time, and has decreased to only about a tenth  of its maximum value  for the largest values of $\tau$ used in the simulations, even for the very moderate values of $g$ and side-branch length.}

 \begin{figure}[h]
    \centering   \includegraphics [width=10cm, height=6cm]{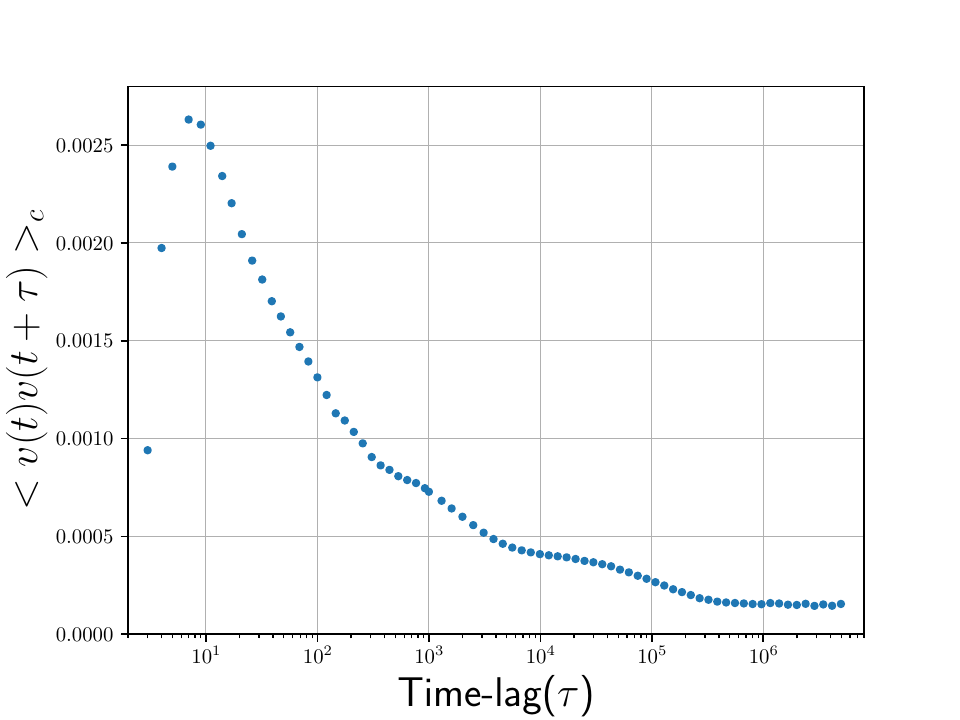}
    \caption{\it Graph showing the velocity-velocity autocorrelation function as a function of time lag ($\tau$) on a regular comb at $g=0.5$. The x-scale is logarithmic. The autocorrelation function is calculated using the velocity at each timestep for $\tau < 1000$, and the velocity averaged over 100 timesteps for calculating the autocorrelation function at $\tau >1000$.}
    \label{fig:enter-label}
\end{figure}

\subsection*{D. Trajectories of mobile and immobile particles}

In Fig. 1 in the main text, we show the percolation cluster with particles traversing through it in the presence of a bias. We colored particles with either red or black, denoting either mobile or trapped particles respectively. Below (figure 4), we show the trajectory of one particle from each class. Mobile particles contribute to the right peak in the velocity plot (Fig. 8) in the main text.

 \begin{figure} [h]
    \centering   \includegraphics [width=7cm, height=7cm]{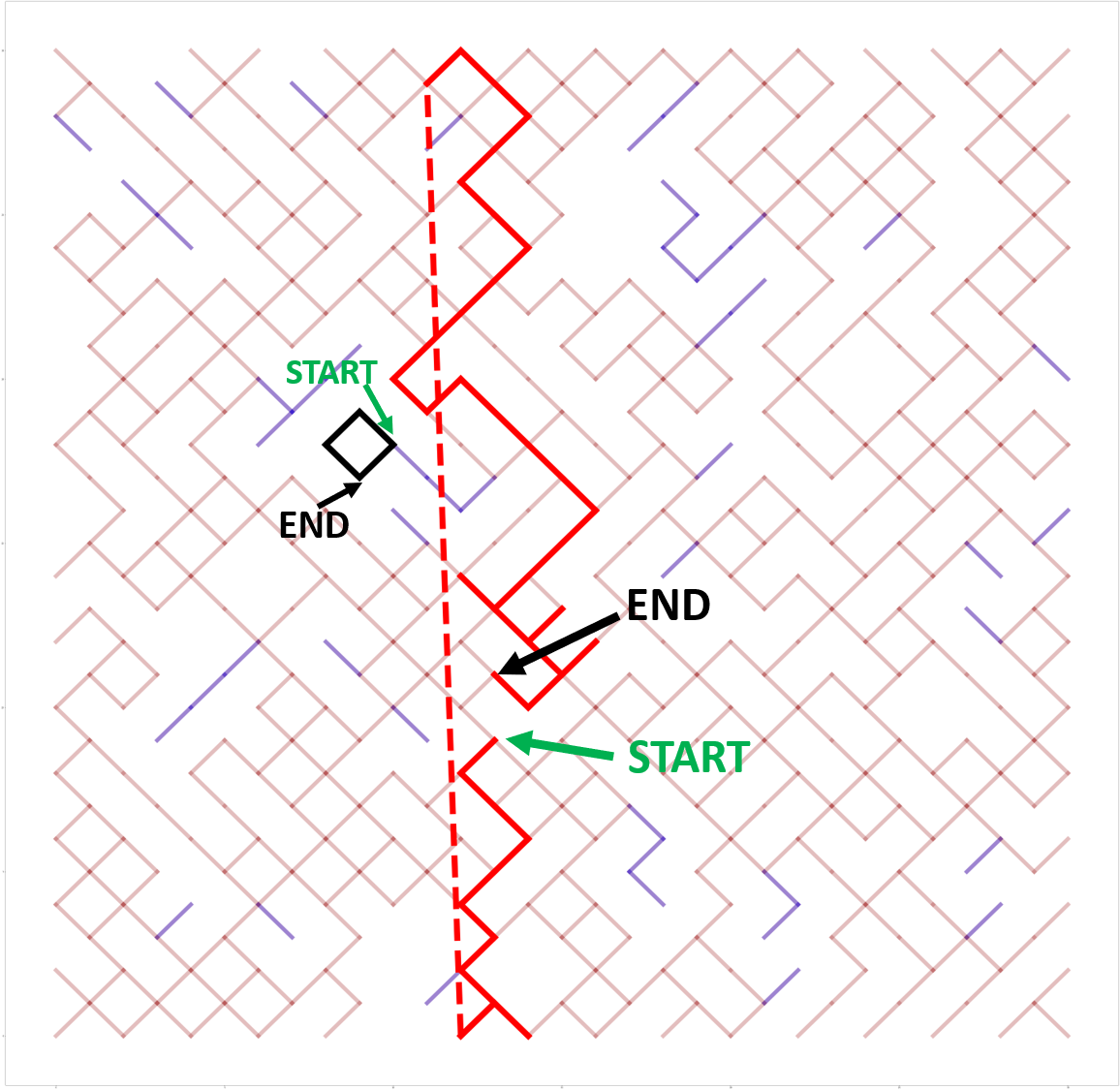}
    \caption{\it Plot showing the trajectory of a red and a black particle, over the course of their evolution for $T=100$ timesteps on the percolation cluster as considered in Fig. 1 in the main text.}
    \label{fig:enter-label}
\end{figure}

\newpage

\subsection*{E. Behaviour in presence of strong fields}

\paragraph*{}Even when $p > p_{d}$, if the density is too low, for $g=1$ there is no transport: all the particles eventually get stuck in dead-ends of branches from which there is no escape. Even if all the dead-end branches are filled by particles, backbends in the undirected backbone act as traps, and collect particles.   If paths with backbends are eliminated, macroscopic transport in the steady state is possible only for $\rho > \rho_{c}(p)$, where $\rho_c(p)$ is the fractional number of sites in the backbone that are in the downward segments of branches. \textcolor{black}{If we specify the current in an open system, the particle density adjusts and the steady state has a very different behaviour [Ramaswamy and Barma 1987].}

\paragraph*{}If $p > p_{d}$, and $\rho  < \rho_c(p)$, with $(1-g)$ non-zero but small, there is a small current, as particles can go over potential barriers. Different traps have different barrier heights. In time, the traps with smaller barrier heights lose particles \textcolor{black}{on average}, and traps with greater heights gain them, so that the effective depths of all traps tend  to equalize. In the long-time steady state, for large bias, the barrier heights of almost all traps have only two values:  $r$ or $(r+1)$, where $r$ is a non-negative integer, whose value is determined by the overall density. Then, the activated current will vary as $(1-g)^{r}$,  as g tends to 1.  The value of this integer $r$, defined in the limit of $g$ tending to 1, will  show jump discontinuities as $\rho$ is changed.



\begin{thebibliography}{0}%
\makeatletter
\providecommand \@ifxundefined [1]{%
 \@ifx{#1\undefined}
}%
\providecommand \@ifnum [1]{%
 \ifnum #1\expandafter \@firstoftwo
 \else \expandafter \@secondoftwo
 \fi
}%
\providecommand \@ifx [1]{%
 \ifx #1\expandafter \@firstoftwo
 \else \expandafter \@secondoftwo
 \fi
}%
\providecommand \natexlab [1]{#1}%
\providecommand \enquote  [1]{``#1''}%
\providecommand \bibnamefont  [1]{#1}%
\providecommand \bibfnamefont [1]{#1}%
\providecommand \citenamefont [1]{#1}%
\providecommand \href@noop [0]{\@secondoftwo}%
\providecommand \href [0]{\begingroup \@sanitize@url \@href}%
\providecommand \@href[1]{\@@startlink{#1}\@@href}%
\providecommand \@@href[1]{\endgroup#1\@@endlink}%
\providecommand \@sanitize@url [0]{\catcode `\\12\catcode `\$12\catcode `\&12\catcode `\#12\catcode `\^12\catcode `\_12\catcode `\%12\relax}%
\providecommand \@@startlink[1]{}%
\providecommand \@@endlink[0]{}%
\providecommand \url  [0]{\begingroup\@sanitize@url \@url }%
\providecommand \@url [1]{\endgroup\@href {#1}{\urlprefix }}%
\providecommand \urlprefix  [0]{URL }%
\providecommand \Eprint [0]{\href }%
\providecommand \doibase [0]{https://doi.org/}%
\providecommand \selectlanguage [0]{\@gobble}%
\providecommand \bibinfo  [0]{\@secondoftwo}%
\providecommand \bibfield  [0]{\@secondoftwo}%
\providecommand \translation [1]{[#1]}%
\providecommand \BibitemOpen [0]{}%
\providecommand \bibitemStop [0]{}%
\providecommand \bibitemNoStop [0]{.\EOS\space}%
\providecommand \EOS [0]{\spacefactor3000\relax}%
\providecommand \BibitemShut  [1]{\csname bibitem#1\endcsname}%
\let\auto@bib@innerbib\@empty
\end{thebibliography}%


\begin{thebibliography}{99}
\bibitem{franklin}  https://profiles.nlm.nih.gov/spotlight/kr/feature/ coal; The Holes in Coal: Research at BCURA and in Paris, 1942-51

\bibitem{broadbent-hammersley} Broadbent, S. and Hammersley, J. (1957) Percolation Processes I. Crystals and Mazes. Proceedings of the Cambridge Philosophical Society, 53, 629-641. 

\bibitem{applications} L.v. Fischer, A.H.v. Gordon, An Introduction to Gel Chromatography, North-Holland, Amsterdam, 1969, p. 338;

\bibitem{devos}  J. De Vos, K. Broeckhoven, S. Eeltink, Advances in ultrahigh-pressure liquid chromatography technology and system design, Anal. Chem. 88 (1) (2016) 262–278; 

\bibitem{sahimi}  M. Sahimi, Flow phenomena in rocks: from continuum models to fractals, percolation, cellular automata, and simulated annealing, Reviews of modern physics 65, 1393 (1993).

\bibitem{biassed-diffusion-3} H. Böttger, V.V. Bryksin, Hopping conductivity in ordered and disordered systems (III), Phys. Status Solidi (B) 113 (1) (1982) 9–49;


\bibitem{de-gennes} P. G. de Gennes et al., La percolation: un concept unificateur, La recherche 7, 919 (1976).
\bibitem{biassed-diffusion-1} S. Havlin and D. Ben-Avraham, Diffusion in disordered media, Advances in physics 36, 695 (1987).

\bibitem{biassed-diffusion-2} Bouchaud, J.-P., \& Georges, A. (1990). Anomalous diffusion in disordered media: Statistical mechanisms, models and physical applications. Physics Reports, 195(4-5), 127–293. 

\bibitem{ramaswamy-barma}  R Ramaswamy and M Barma, Transport in random networks in a field: interacting particles,   1987 J. Phys. A: Math. Gen. 20 2973.

\bibitem{derrida} B. Derrida, An exactly soluble non-equilibrium system : The
asymmetric simple exclusion process,  Physics Reports 301 (1998) 65—83.
\bibitem{mallick}   Olivier Golinelli and Kirone Mallick,  The asymmetric simple exclusion process: an integrable model for non-equilibrium statistical mechanics, 2006 J. Phys. A: Math. Gen. 39 12679
\bibitem{liggett} T. Liggett, Ergodic theorems for the asymmetric simple exclusion process,   Trans. Amer. Math. Soc. 213 (1975), 237-261.
\bibitem{barma-dhar}  Barma M and Dhar D, Directed diffusion in a percolation network, 1983 J. Phys. C: Solid State Phys. {\bf 16} 1451; 


\bibitem{dhar-stauffer}  D. Dhar, D. Stauffer, Drift and trapping in biased diffusion on disordered lattices, Internat. J. Modern Phys. C 09 (02) (1998) 349–355.

\bibitem{ohtsuki} T. Ohtsuki and T. Keyes, Mobility and Linear Response Theory on Percolation Lattices, Phys. Rev. Lett. (1984) 52, 1177. 

\bibitem{supplemental} See Supplemental Material.

\bibitem{persistence} Satya N. Majumdar, Persistence in non-equilibrium systems, Current Science, {\bf 77} (1999) 370-375.


\bibitem{random-comb-1}  S. R. White and M. Barma, Field-induced drift and trapping in percolation networks, J Phys. A: Mathematical and General 17, 2995 (1984); 

\bibitem{random-comb-2} S. B. Yuste, E. Abad, and A. Baumgaertner, Anomalous diffusion and dynamics of fluorescence recovery after photobleaching in the random-comb model, Phys. Rev. E 94, 012118 (2016); 

\bibitem{random-comb-3} V. M\'endez and A. Iomin, Comb-like models for transport along spiny dendrites, Chaos, Solitons \& Fractals 53, 46 (2013);

\bibitem{random-comb-4} G\'erard Ben Arous, Alexander Fribergh, Nina Gantert, and Alan Hammond. Biased random walks on Galton-Watson trees with leaves. Ann. Probab., 40(1):280–338, 2012;

\bibitem{random-comb-5} J. D. Kotak and M. Barma, Bias induced drift and trapping on random combs and the bethe lattice: Fluctuation regime and first order phase transitions, Physica A: Statistical Mechanics and its Applications, 127311 (2022);

\bibitem{random-comb-6} N. Pottier, Diffusion on random comblike structures: field-induced trapping effects, Physica A: Statistical Mechanics and its Applications 216, 1 (1995);

\bibitem{random-comb-7}  V. Balakrishnan and C. Van den Broeck, Transport properties on a random comb, Physica A: Statistical Mechanics and its Applications 217, 1 (1995). 


\bibitem{heterogeneity1} Hans Sillescu, Heterogeneity at the glass transition: a review, Journal of Non-Crystalline Solids, Volume 243, Issues 2–3, Pages 81-108 (1999). 

\bibitem{heterogeneity2} Kallol Paul, Anoop Mutneja, Saroj Kumar Nandi and Smarajit Karmakar, Dynamical heterogeneity in active glasses is inherently different from its equilibrium behavior, Proc. National Acad. Sciences (USA), (2023) 120 (34) e2217073120. 

\bibitem{many-body-localization-1} F. Alet, N. Laflorencie, Many-body localization: An introduction and selected topics, 
Comptes Rendus Physique, {\bf  19}, Issue 6, (2018) , Pages 498-525.
\bibitem{many-body-localization-2} Yu. Kagan and L. A. Maksimov, Localization in a system of interacting particles diffusing in a regular crystal, Zh. Eksp. Teor. Fiz. 87, 348-365 (July 1984). 

\bibitem{pradhan-dd} Punyabrata Pradhan  and Deepak Dhar, Probability distribution of residence times of grains in models of rice piles, Phys. Rev. E 73, (2006) 021303. 


\end{thebibliography}
\end{document}